# EXTENDING RECORD AND PLAYBACK TECHNOLOGIES TO SUPPORT COOPERATIVE LEARNING

*Esha Ray[1], Ken Watts[2], Rick Adrion[3]*

*Abstract* - We have long-term experience with developing and employing multimedia materials for on-campus and distance education. We also are assessing the efficacy of cooperative learning where groups of learners explore, with guidance from an instructor, the learning environment and construct models of meaning based on their shared learning experiences. Our core technologies capture and store classroom events, but are record-and-playback technologies focused on delivering content to individual learners. We describe an extension of our technology, Cooperative Learning in MANIC (CLIMANIC), which allows groups of learners and teachers to collaborate and communicate. We describe our current assessment of CLIMANIC and future plans for more extensive evaluation.

*Index Terms* – Cooperative Learning, Collaboration, Distance Learning, Learning Technologies

## MOTIVATION

Within the Research in Presentation Production for Learning Electronically (RIPPLES) Group at the University of Massachusetts Amherst, we have long-term experience in developing and employing multimedia materials for on-campus and distance education. We capture both live and authored presentations, couple them with text, graphics and search/index mechanisms, and deliver the content though streaming servers and on CD/DVD [1,2,3] using the Multimedia Asynchronous Networked Individualized Courseware (MANIC) framework. The MANIC project had its serendipitous beginnings in 1996, arising out of an unexpected confluence of interests among researchers in the areas of multimedia systems and education. The RIPPLES Group developed dozens of full-semester courses that are currently being delivered to students using the MANIC technology. Students include those enrolled in University of Massachusetts Amherst distance education courses, as well as on-campus students at a number of other universities, including Polytechnic University (Westchester campus), five campuses of North Carolina State University system and, soon, several state and community colleges in Massachusetts and the City University of New York.

In this paper, we take a constructivist view of learning, where learners ascribe their own meaning to the world by constructing understanding through experimentation and discovery. *Cooperative learning* is a constructivist-oriented instructional model, and we view constructivist cooperative learning as an environment (physical or virtual) where groups of learners explore (with guidance from an instructor) the learning environment and construct, elaborate, scaffold and adapt their models of meaning based on their shared learning experiences and active investigation of provided materials. Learning is measured by the evaluation of on-going reporting by the learners of their projects and experiences. We envision a system for creating and delivering courseware that will include mechanisms for discovery and constructing knowledge and for collaboration to support cooperative learning. Such a facility would effectively create a virtual classroom experience that can be shared by geographically separated students while providing individualized experiences for identical content units.

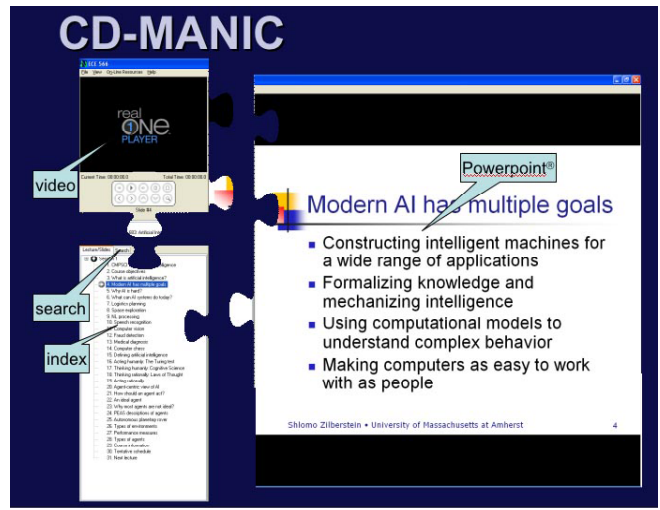

FIGURE 1. CD MANIC COMPONENTS

Most systems, such as the original MANIC approach, that capture and store classroom events are *record-and-playback* technologies aimed at producing courseware for distance learning and for review/documentation for on-campus students. These systems generally are passive media, focused on delivering content to individual learners. They have limited ability to support learner navigation, notation and collaboration, keys to knowledge discovery and construction. While we have shown MANIC to be effective in support of constructivist approaches [4], record-and-playback technologies are not inherently constructivist [5] and clearly lack mechanisms for collaboration.

---
[1] Esha Ray, Research Assistant, University of Massachusetts Amherst, RIPPLES Laboratory, esha@cs.umass.edu
[2] Ken Watts, Senior Software Engineer, University of Massachusetts Amherst, RIPPLES Laboratory, watts@cs.umass.edu
[3] W. Richards Adrion, Professor of Computer science, University of Massachusetts Amherst, RIPPLES Laboratory, adrion@cs.umass.edu

Collaboration systems that seamlessly connect geographically dispersed participants hold tremendous potential for many applications, but especially distance learning. Collaboration mechanisms like Lotus Sametime [6], ConferenceXP [7], NetMeeting [8], and Application Sharing over MSN Messenger [18] have already proven to be efficient approaches to improve contact and group meetings among geographically dispersed users. We expected that such application-sharing products could be employed to add a collaboration mechanism to MANIC courseware, but a number of technical issues arose.

Existing collaboration tools and streaming media players (e.g., Real player and Windows Media Player) do not adequately support collaborative video viewing. If multiple learners use NetMeeting, for example, to establish a common "call" to a MANIC application, any one of the participants should be able to start a lecture and share viewing with others. While the remote participants can see and control the entire application, the video is of low quality and jittery. This is due to the fact that the MANIC video data stream is decoded and then delivered to the participant who runs the MANIC application. This stream is then encoded using the application sharing system algorithms and sent to the other participants' machines. As a result, performance is poor and a large amount of network bandwidth is consumed. None of the application sharing products performs well in this scenario. Integrating one of these application-sharing mechanisms into MANIC typically would limit use to Windows-based systems. Since we wish to support MANIC on multiple platforms, constructing an independent collaboration environment that supports synchronous collaboration activities is the key to building MANIC from an individualized to a truly collaborative system. Such a collaborative environment will require two key components.

- A distributed MANIC version that can be viewed among participants with shared controls.
- A communication system for discussions among the learners and teachers.

We introduce the notion of a "session" where every participant in the session has a separate copy of the same MANIC courseware running on their computer but the playback is a shared experience among all the users. Students can communicate via chat, share control of the session, and hence learn in a collaborative environment.

## RELATED WORK

Recent studies [9,10] have shown cooperative learning to be effective in computer science education and several other studies [11,12] report that underrepresented groups and women learn better when these learning approaches are applied effectively. Our intention is to engage students as active participants in a rich problem solving and cooperative learning environment and to facilitate cooperative learning on-campus, between the institutions and with (a)synchronous distance learners. Contact among the students is an important and crucial factor to guarantee the success of MANIC in a cooperative learning context. Off-campus students lack the environment of a real classroom where they can communicate and cooperate easily with instructors and other students. This lack of contact is a hindrance to successful collaborative/group work. In current versions of MANIC we support asynchronous collaboration in the form of threaded discussions and note taking activities. But due to the lack of necessary synchronous collaboration support, the students have not known nor have been in contact with other (on- or off-campus) class members; they have not interacted with the instructors beyond email and hence have not been actively engaged in the learning process.

A number of systems [13,14] provide note-taking mechanisms. The CoNote [19] System supports annotations on a set of documents shared among a group. The RIPPLES group added a note-taking and note-sharing extension that allows students to record and view personal notes (text or simple drawings) linked to specific slides in CD-MANIC courseware. The Microsoft Research Annotation System (MRAS) is a prototype system for annotating multimedia presentations that provides a window for annotations linked to the current video/slide state. MRAS has also been studied as a tool for collaboration.

Recently, a number of collaboration tools have been introduced to support collaboration ranging from individual users to large enterprises. Lotus Instant Messaging and Web Conferencing (Sametime) was designed to facilitate communication among coworkers in different geographic locations. It is a synchronous groupware application and comes with its own directory system for authentication, and can be "plugged into" other Lightweight Directory Access Protocol (LDAP) compliant directory services. Communication is by instant messaging or video conferencing over IP. Sametime allows sharing control of such applications as word processors, spreadsheet applications, presentation applications, and whiteboards.

Microsoft's NetMeeting and the recent Conference XP [7] include many of the same features as Lotus Sametime. ConferenceXP, designed to take advantage of recent advances in technology such as the Internet2 Abilene network, is a platform for designing and implementing high-end interactive collaborative applications. ConferenceXP integrates audio, video, and network technologies into an environment for synchronous collaboration activities. It also supports online forums that connect remote users. One of its interesting features is that it provides the option of a variety of distance collaboration applications, such as one-to-one messaging, one-to-many instructional presentations for distance learning, and interactive group-to-group remote collaborative conferences and discussions. Microsoft claims that ConferenceXP reduces network bandwidth issues that have been a bottleneck in the building of successful collaboration systems; i.e., "it utilizes highly optimized audio and video codecs included with Windows to minimize bandwidth and CPU utilization [and] the ConferenceXP network transport dynamically adjusts to handle poor network conditions and low bandwidth situations" [7].

The University of Hannover, Germany and the Free University of Bozen in Italy [15] employ Web-based Distributed Authoring and Versioning (WebDAV) to implement a system that supports asynchronous collaboration activities and use IBM Sametime to provide support for synchronous collaboration activities. Both the asynchronous and synchronous collaboration sessions share a user management mechanism that is implemented based on a LDAP server. The server is able to manage users and user groups on the basis of secure authentication/authorization. This system employs a full Web interface such that both the asynchronous and synchronous collaboration sessions have the facility to be started through standard Web browsers.

The Microsoft group that developed MRAS has developed TELEP [20], a system that supports shared viewing and control of lectures using NetMeeting. TELEP allowed distributed individuals to collectively watch video using shared VCR controls (see Figure 2). While adapting the TELEP approach for MANIC seemed promising, our goals were different. With TELEP, they hoped to incorporate remote viewers into a "live" classroom. We plan some experimentation with a similar approach later (see Future Work), but our goal is to support cooperative group learning, particularly with existing MANIC courseware. Moreover, the TELEP project seems to have been set aside for other priorities before a fully integrated system was developed. The TELEP Project focused (naturally) on proprietary Microsoft platforms and we plan a more open-systems approach.

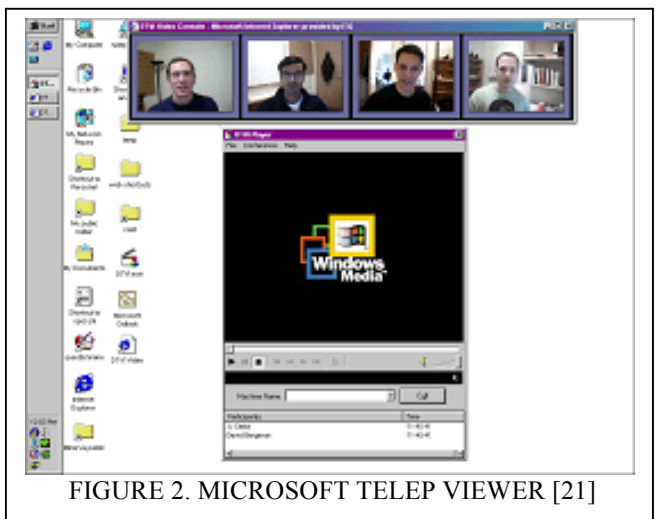

FIGURE 2. MICROSOFT TELEP VIEWER [21]

## CLIMANIC

A key component of the enhanced version of MANIC, Cooperative Learning In MANIC (CLIMANIC), will be the ability to collaboratively view the courseware. Its purpose is to approximate, as closely as possible, the experience of face-to-face group meetings amongst both off and on-campus students and instructors. For example, if one participant in a group presses the play button, the video should play for all the other remote participants. This means that the interaction of all the students in the group with respect to MANIC should be synchronized.

To solve the above problem of video streaming we need partial sharing of the MANIC application. Since we want the video to be played independently on each computer, participants run a separate copy of the courseware on their respective computers. Therefore, instead of sharing the entire application remotely with all other participants we just share the controls of the application amongst all the participants, which minimizes constraints on bandwidth.

CLIMANIC provides an efficient team awareness support for all synchronous collaboration activities based on a centralized database using mySQL as the database server. The database contains the information for all of the courses being offered, the group IDs of all the groups, whether a group is active or not, the IP addresses of all of the members in various groups, and the participant who currently has the control of the group. Any participant who wants to enter a study group will be required to connect to the database and query for the active groups. Active groups are the groups that are still waiting for more participants. The participant then queries the database for the IP address of the participant who has the control of the group, the group leader. The participant then connects to the group leader.

The group leader sends a message to all of the other participants that a new participant has entered the group. A participant can also form a new group and wait for others to join. After all of the group members decide that they don't want any more participants to enter, they start a session and the group becomes inactive. By default, the first participant is the one who has the control of the session and hence is the group leader. If the group leader wants to leave a session, anyone else can be asked to take the control. If the leader leaves abruptly, such as being logged off due to network problems, the system automatically hands the control to the participant who had joined earliest among the existing participants of the group. During a session, any other participant has the option of asking for the control of the session and becoming the group leader. Since participants are able to request control from the group leader, this can lead to multiple participants requesting for the control at the same time. In such a scenario the group leader will see multiple messages like "User X has asked for the control" and will have the choice to accept or decline the request. Once the leader accepts a request, all of the other requests disappear. When a request is declined, an appropriate message is sent to that participant.

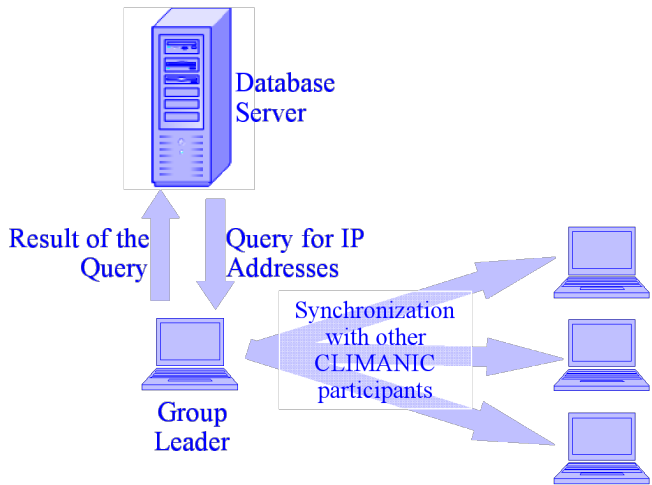

FIGURE 3. PEER TO PEER INTERACTIONS

A participant will have the option to logout and exit a group. The entry for that particular participant will then be removed from the database. In a scenario where the participants who have been logged out want to re-join the group, the group leader will have the choice to make the group active so that those participants can reenter. The database will then be updated with the new information.

The system uses a peer-to-peer P2P) approach to synchronize the courseware for all of the participants. When a participant takes or has the control of the session, he queries the database for the IP addresses of all of the other participants. During a session, the view of the MANIC courseware of the current leader is a shared experience among all of the participants. For example, if the leader clicks to move the lecture slide forward from slide 10 to slide 11, a message is sent to all the other clients (separate MANIC applications running on the computers of other participants) to update to slide 11. Hence, everybody's view is synchronized (see Figure 3).

CLIMANIC is based on the existing C++ MANIC (core-MANIC) implementation together with a Java implementation of the collaboration mechanisms and the Java Native Interface (JNI) technology [16] that provides an interface between the collaboration mechanisms and core-MANIC. Since the implementation of different modules will be in different programming languages, we had to implement an interface using JNI technology. The JNI interface is basically a way through which the core-MANIC code interacts with the collaboration mechanisms and vice versa. The JNI connector provides a messaging service between the different modules of the system by passing and converting data from one module such that the modules can communicate. This interface will be the *middleware* that enables the two systems to talk (see Figure 4).

Any action on the leader's part that changes the view of a CLIMANIC application generates an event that sends a message to the JNI interface. The CLIMANIC collaboration mechanisms receive the message when an event is generated from core-MANIC, parses it and then calls the appropriate function to perform the task, for example, to synchronize the courseware of all of the other participants to pause the video. In this scenario, the function will send a message to all of the other clients through the JNI interface.

There has to be an effective way for the participants to interact. The system requires a communication channel for its users. Text-based chat and instant messaging can consist of two (or more) participants exchanging instant messages. Initially, we integrated AOL Instant Messenger, which requires participants to register for an AOL screen name. (Later, we may move to other conference/messaging tools.) The integration of AIM with the collaborative system will be done using a JAIM [17], a Java library that implements the AOL TOC protocol.

The current design of CLIMANIC combines centralized and P2P approaches to group communication. All participants query a centralized database to submit and obtain IP address information used for P2P communication between various CLIMANIC clients during a session.

## ASSESSMENT

To provide an initial evaluation of CLIMANIC, we are creating courseware for the NSF Engineering Research Center for Collaborative Adaptive Sensing of the Atmosphere (CASA). CASA is comprised of four collaborating universities - the University of Massachusetts

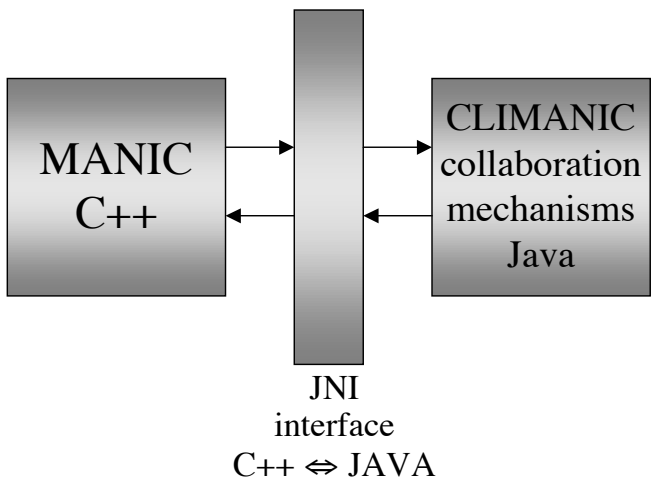

FIGURE 4. SYSTEM INTERACTIONS

Amherst, the University of Oklahoma, Colorado State University, and the University of Puerto Rico. Additional participants include: University of Delaware, Mount Holyoke College, Rice University, and University of Colorado Boulder, Raytheon, IBM, M/A-Com, Telephonics, Flash Technology, Vaisala, Baron Services, The Weather Channel, Weather Services International, Vieux Associates, and the Texas Medical Center. This $40 million center brings together a multidisciplinary group of engineers, computer scientists, meteorologists, sociologists, graduate and undergraduate students, and industry and government representatives to conduct fundamental research, develop enabling technology, and deploy prototype engineering systems based on a new paradigm: Distributed Collaborative Adaptive Sensing (DCAS). CASA represents a unique opportunity to evaluatethe CLIMANIC in a context where geographically dispersed collaborators from diverse institutions (universities, industry, government) and from a variety of backgrounds (faculty, students, staff, administrators, emergency response personnel, broadcasters, etc.) need to obtain a working knowledge of the many disciplines (networking, radar, sensors, weather modeling, etc.) involved in CASA.

We are beginning by creating courseware authored by the primary research leadership of CASA and targeted for the immediate CASA faculty, staff and graduate students. As a second stage, we will distribute the courseware to undergraduate students participating in the CASA NSF Research Experiences for Undergraduates (REU) Program beginning in Summer 2004. In Fall 2004, we plan a wider distribution to all CASA participants and later to high schools as part of the CASA outreach program. The CASA Education and Outreach Program is co-directed by RIPPLES researcher, Professor Wayne Burleson.

Our primary tool for assessing CLIMANIC is the Learner Logger [1]. Over the last few years, we have captured a substantial amount of data using the Learner Logger on our current RIPPLES technologies in traditional distance learning and on-campus settings. From the patterns of events, usage and navigation, we have gained valuable insight into whether learners follow a traditional objectivist (sequential) path through the material or what can be inferred to be a more constructivist path: multiple entry points, non-linear navigation, use of search to access course and external materials, etc. The Learner Logger will be extended to capture data on CLIMANIC sessions, allowing us to conduct experiments to study various aspects of cooperative group learning, e.g., group interactions, group dynamics and group restructuring.

In the current experiment with the CASA core faculty, staff and graduate students, we are using qualitative and quantitative questionnaires, one-on-one interviews, focus groups and observations, augmented by the Learner Logger embedded assessment tools. Preliminary data available in late Spring 2004 will be used to guide the experiment with REU students over Summer 2004. REU students will work in groups. Some course materials will be assigned to supplement the CLIMANIC courseware, while others will be "discovered" while students are engaged in assigned problems and/or projects. Performance will be assessed using constructivist assessment techniques (written and oral, individual and group reports, assignments) supported by questionnaires, interviews, student observations, pre and post testing and Learner Logger data.

## FUTURE PLANS

We will employ a cooperative learning approach and deploy CLIMANIC, assessing the use of educational pedagogies and technologies in two more experiments in Fall 2002.2004. These include:

- Introductory UG classes at City College of New York (CCNY): use of technology-supported, active, peer and cooperative learning in large, diverse introductory courses with a special emphasis on retention of women and underrepresented minorities.
- Upper-level UG class taught on-campus and via distance education at University of Massachusetts Amherst (UMass)**:** integration of on-campus and at-a-distance students in a problem-focused, cooperative learning model of instructional delivery.

We chose the CCNY experiment for several reasons. CCNY is very concerned about its retention and graduation rates. CCNY is a minority-serving institution with under-represented minorities in 2002 comprising 66% of undergraduate degrees in engineering and 33% in computer science. Almost 20% of CCNY engineering undergraduates are women. Positive results will have significant national impact by identifying a real opportunity to increase the diversity of CS graduates.

We chose the UMass experiment because the Computer Science Department offers almost all of its 500-level (advanced undergraduate) courses through Professional Education for Engineering and Applied Science (PEEAS) program and the National Technical University (NTU). Distance education students are treated quite independently of on-campus students at present and require additional instructor effort. Off-campus students have the flexibility to "learn anytime, anywhere," but they often fall behind their on-campus peers. However, many of the distance learners in computer science are practicing engineers and can bring considerable real-world experience to a team with on-campus students. Integrating on- and off-campus students will benefit both. Success in this experiment can serve as a national model for integrating distance and traditional education.

## CONCLUSIONS

In this paper we presented the design of the CLIMANIC system that we have implemented to efficiently extend

MANIC to include a collaborative environment for cooperative learning. Our goal for CLIMANIC was to create a flexible and extensible infrastructure supporting interactive collaborative environments. We chose an approach that employs both centralized and P2P techniques to offer a sound and robust infrastructure.

As Murphy [10] notes, "There is a wealth of education research confirming the benefits of active and cooperative learning in higher education; benefits which include a deeper understanding of the material, higher academic achievement, increased student persistence and retention, positive peer relationships and higher self esteem." However, there are relatively few empirical studies of constructivism in computer science education and few have been undertaken in large public institutions. The data we collect should have broad impact on the adoption of cooperative learning models and on the design of learning technologies to support them.

## ACKNOWLEDGMENTS

We wish to thank the other members of the RIPPLES Group: Wayne Burleson, Jim Kurose, Wendy Cooper, Dula Kumela, James Cori, and Byron Wallace for their assistance and support.

This work was partially supported by the National Science Foundation under EIA- 9979833.